\documentclass[onecolumn]{autart}    % Enable this line and disable the
\usepackage{graphicx}          % Include this line if your
\usepackage{makeidx,graphics,subfigure}
\usepackage{graphicx}% Include figure files
\usepackage{dcolumn}% Align table columns on decimal point
\usepackage{bm}% bold math
\usepackage{multicol}
\usepackage{amsmath} % assumes amsmath package installed
\usepackage{amssymb}  % assumes amsmath package installed

\def\vec#1{\mathbf{#1}}

\begin{document}

\begin{frontmatter}

\title{Set Stabilizability of Quantum Systems\thanksref{footnoteinfo}} % Title, preferably not more

\thanks[footnoteinfo]{This paper was not presented at any IFAC
meeting. Corresponding author M.Zhang. Tel. +0086-731-84573333. Fax
+0086-731-84573335}

\author[C1]{Ming Zhang}\ead{zhangming@nudt.edu.cn},    % Add the
%\ead{hydai@nudt.edu.cn},               % e-mail address
\author[C2]{Zairong Xi},
%\ead{zrxi@iss.ac.cn}  % (ead) as shown
\author[C3]{Tzyh-Jong Tarn}

\address[C1]{Department of Automatic Control,
College of Mechatronics and Automation, National University of
Defense Technology, Changsha, Hunan 410073,  People's
Republic of China}  % Please supply
\address[C2]{Key Laboratory of
Systems and Control, Chinese Academy of Sciences, Beijing, 100080,
 People's Republic of China}        % here.
 \address[C3]{Department of Electrical and Systems Engineering, Washington University in St. Louis, M
O, 63130-4899}        % here.

                             % keyword list or with the
                                          % help of the Automatica
                                          % keyword wizard

\begin{abstract}                          % Abstract of not more than 200 words.
 We explore set-stabilizability  by constrained controls,  and  both controllability and stabilizability can be regarded as the special case of set-stabilizability.   We not only clarify how to define  equilibrium points of Schr$\ddot{o}$dinger Equations, but also establish the necessary and sufficient conditions  for stabilizability of quantum systems.  Unfortunately, it is revealed  that the necessary conditions  are quite strict for stabilizability of some concrete quantum systems like nuclear spin systems, and this further justifies  the introduction of  set-stabilizability notion.  It is also exemplified  that set-stabilizability  can  be used for investigating quantum information processing problems including quantum information storage and entangled states generation.
\end{abstract}

\end{frontmatter}

\section{Motivation and Introduction}
The concepts of controllability and observability are important  contribution of control theorists to the science, technology, and engineering domain.
 With the introduction of these structural concepts, we begin to deeply understand the relationship between the input-output description and state-space description. The concept of  controllability was first  proposed for linear systems by R. E.  Kalman in his remarkable paper\cite{Kalman} in 1960s.
Controllability of nonlinear systems was further investigated by H. J. Sussmann and V. Jurdjevic\cite{Sussmann} and R. Hermann and A. J. Krener\cite{Hermann} in 1970s.

 Quantum control
theory has been developed ever since last century\cite{Belavkin,Huang2,Ong
3,Clark4}. Recently, quantum information and quantum computation is the
focus of reseach\cite{quantum information}. A great progress has
been made in the domain of  quantum control\cite{quantum
control,dp}, in which the controllability of quantum systems is a
fundamental issue.  In 1980s, controllability of quantum systems was first explored by  G. M.  Huang,  T.-J. Tarn and J. W. Clark\cite{Huang2}. The different notations of controllability have
been exploited in \cite{R95,W06,S01,Z05,A03,W07,T01}. Specially, the
controllability of quantum open systems has been studied by some
researchers\cite{Schirmer 2003-19,Altafini 04-21,Lloyd 00-22,zm 06}.

To manipulate a quantum system, it is not sufficient to just know whether or not the quantum system is controllable.
 It is necessary to know how to construct permissible controls to steer the quantum system within the the specified time $T_s$ in some applications. We need  explore controllability of quantum systems under different permissible control conditions such as bounded controls and  time continuous function controls, and exploit the impact of  admissible control conditions on the performance indices including transition time.

 Given a control system, whether quantum or classical, the first and most important question about its various properties is to investigate whether or not it is stable. The most useful and general approach for studying the stability and stabilizability is the theory introduced by Russian mathematician A. M. Lyapunov\cite{Ly} in the late 19$^{th}$ century. Lyapunov's pioneering work on stability received little attention outside Russia,
although it was translated into French in 1908 (at the instigation of Poincare), and
reprinted by Princeton University Press in 1947. The publication of the work of Lure
and a book\cite{Ls} by La Salle and Lefschetz brought Lyapunov's work to the attention of the
larger control engineering community in the early 1960's. Several quantitative stability concepts like finite-time stability\cite{fs}, Lipschitz stability\cite{ls}, partial stablity\cite{partial} and practical stability\cite{Ls} had been investigated based on Lyapunov's great work in 20$^{th}$ century. Set stability of dynamical systems was specifically discussed by Heinen James Albin\cite{Albin} in 1969. In 2002, S. K. Phooi et al.\cite{Phooi} further proposed the broad-sense Lyapunov function and generalized the notion of stability in the sense of Lyapunov.

From control theory point of view, we not only need to investigate whether a dynamical system is stable or not, but also need to explore whether or not a controlled dynamical system is stabilizable by permissible control.

   In recent applications like quantum information storage, one of the important questions is  whether or not a given state of  quantum systems can be stabilizable by permissible controls. To study the stabilizability problem of quantum systems, we will  have to exploit how to define the equilibrium points of Schr$\ddot{o}$dinger Equation. Just by investigating stabilizability problem of controlled quantum closed system, we gradually realize that the physical conditions for stabilizability of quantum systems are too strict in some concrete  systems like nuclear-soin systems.  These observations  indicate that we need to weaken a more general framework.

  To overcome the aforementioned difficulties, we try to generalize the concepts of both controllability and stabilizability. With quantum control problems in mind, we will propose a new notion of set-stabilizability: given a pair of quantum state sets $S_0$ and $S_1$, a quantum system is  $S_1$-stabilizability   from $S_0$ within the specified time $T_s$ under constrained  control conditions if,  for any initial state in $S_0$, on can always find  permissible  controls to steer the quantum system from an arbitrary initial state in $S_0$ to another arbitrary  state in the set $S_1$ in the finite time $t_f$ with $0<t_f\leq{T_s}$, and to further keep the system state  stay in the set $S_1$  when  $t\geq{T_s}$.

 It is interesting to underline the following observations: (1) When $S_1$ is a one-point set, set-stabilizability notation is reduced to stabilizability.
 (2) When both $S_0$ and $S_1$ are the state space itself and unconstrained controls are permitted, the concept of set stabilizability is reduced to controllability proposed by Kalman\cite{Kalman}.

However, the set-stabilizability  is not proposed to generalize the concepts of both controllability and stabilizability just for the sake of generalization, without proper motivations. This work can be regarded as one of explorations made by many researchers who hope to investigate what kind of control goal is achievable for quantum systems by various feedback control\cite{JR1992,ZR2003,WM2010,ADL2002,WD2005,HSM2005,MH2007,DP2009,ZWLT2010,JL2001,DJJ2001,QG2010,CXP2008,XYS2006,YK2005}. Stabilization of open quantum systems has been  studied by  \cite{SW2010,QPG2013}. In this research, we  exploit what kind of stabilizability can be expected for quantum closed systems by constrained open-loop controls, and we  would like to emphasize that set stabilizability is attained by coherent control even when stabilizability itself is not an achievable control goal. It should be also underlined that this research is very different from set value analysis\cite{Aubin} and set dynamics\cite{Hong}.

 The rest of this paper are organized as follows. In Sect. II, the concept and properties of set-stabilizability are presented for general dynamical systems, and it is revealed that both controllability and stabilizability are the special case of  set stabilizability.  The notation of set stabilizability and stabilizability are specifically discussed for quantum closed systems. In Sect. III, the set stabilizability and stabilizability problems of single-qubit systems are explored under different constrained controls. We present the necessary and sufficient conditions for quantum systems and give some further discussions on the strictness of stabilizability in Sect. IV, and it is also exemplified that set-stabilizability notation can be used for exploring entanglement generation of two-qubit systems  in this section. The paper concludes with Sect. V.

\section{Basic Concept and Basic Lemma}

\subsection{Set stabilizability and its properties}
For the purpose of further discussions, we first recite the axiomatic definition of a dynamical system presented by Kalman in 1960s\cite{Kalman}.

\textbf{Definition 1}.  A dynamical system is a mathematical structure defined by the following axioms:

($D_1$)  There is given a state space $\Sigma$ and a set of values of time $T$ at which the behaviour of the system is defined; $\Sigma$ is a topological space and $T$ is an ordered topological space which is a subset of the real numbers.

($D_2$)  There is given a topological space $\Omega$ of functions of time defined on $T$, which are the admissible inputs to the system.

($D_3$)  For any initial time  $t_0$ in $T$, any initial state $x_0$ in $\Sigma$, and any input $u$ in $\Omega$ defined for $t\geq{t_0}$, the future states of the system are determined by the transition $\varphi:\Omega\times{T}\times{T}\times\Sigma\rightarrow\Sigma$, which is written as $\varphi_{u}(t;t_0,x_0)=x_{t}$. This function is defined only for $t\geq{t_0}$. Moreover, any $t_0\leq{t_1}\leq{t_2}$ in $T$, any $x_0$ in $\Sigma$, and any fixed $u$ in $\Omega$ defined over $[t_0,t_1]\cap{T}$, the following relations hold:

\begin{equation}
\label{0-1}\varphi_{u}(t_0;t_0,x_0)=x_0,
\end{equation}
\begin{equation}
\label{0-2}\varphi_{u}(t_2;t_0,x_0)=\varphi_{u}(t_2;t_1,\varphi_{u}(t_1;t_0,x_0)).
\end{equation}

In addition, the system must be nonanticipatory, i.e., if $u,v\in\Omega$ and $u=v$ on  $[t_0,t_1]\cap{T}$ we have
\begin{equation}
\label{0-3}\varphi_{u}(t;t_0,x_0)=\varphi_{v}(t;t_0,x_0).
\end{equation}

($D_4$) Every output of the system is  a function $\psi:{T}\times\Sigma\rightarrow$reals.

($D_5$) The functions $\varphi$ and $\psi$ are continuous, with respect to the topologies defined for $\Sigma$, ${T}$, and $\Omega$ and the induced product topologies.

In other words, a dynamical system can be described by $\Xi=\{\Sigma,{T},\Omega,\varphi,\psi\}$. For a dynamical system without output, it can be reduced to $\Xi=\{\Sigma,{T},\Omega,\varphi\}$.

Subsequently, we will recite the concepts of controllability and stabilizability in the aforementioned abstract framework of dynamical systems.

\textbf{Definition 2}. The dynamical system $\Xi=\{\Sigma,{T},\Omega,\varphi\}$ is controllable at time  $t_0\in{T}$, if for any pair of initial state $x_0$ and target state $x_1$ in the state space $\Sigma$, there always exist $t_1\in{T}$ with $t_0\leq{t_1}<\infty$ and  admissible control $u\in\Omega$ such that $x_1=\varphi_{u}(t_1;t_0,x_0)$.

\textbf{Definition 3}. Let $x_1\in{\Sigma}$, then the state $x_1$ of the dynamical system  $\Xi=\{\Sigma,{T},\Omega,\varphi\}$ is stabilizable from $\Sigma$ at time $t_0\in{T}$, if  for any  initial state $x_0\in{\Sigma}$, there exist $t_1\in{T}$ with $t_0\leq{t_1}\leq\infty$ and  admissible control $u\in\Omega$ such that $\varphi_{u}(t;t_0,x_0)\equiv{x_1}$ when $t\geq{t_1}$.

Suppose that both $S_0$ and $S_1$ are subsets of the state space $\Sigma$, we will introduce a new concept of set stabilizability as follows.

\textbf{Definition 4}. The dynamical system  $\Xi=\{\Sigma,{T},\Omega,\varphi\}$ is $S_1$-stabilizable from $S_0$ at time  $t_0\in{T}$, if  for any  initial state $x_0\in{S_0}$ and another arbitrary target state $x_1\in{S_1}$, there exist $t_1\in{T}$ with $t_0\leq{t_1}\leq\infty$ and  admissible control $u\in\Omega$ such that $x_1=\varphi_{u}(t_1;t_0,x_0)$ and $\varphi_{u}(t;t_0,x_0)\in{S_1}$ when $t_1\leq{t}\in{T}$.

From the aforementioned definition, we can easily establish the following properties of set stabilizability.

\textbf{Proposition 1} Suppose that $S_0^{'}$, $S_0$ and $S_1$ are subsets of $\Sigma$

($P_1$) Let $S_0^{'}\subseteq{S_0}$,  a dynamical system  $\Xi=\{\Sigma,{T},\Omega,\varphi\}$  is $S_1$-stabilizable from $S_0^{'}$ at time  $t_0\in{T}$ if it  is $S_1$-stabilizable from $S_0$ at time $t_0\in{T}$.

($P_2$) Let $\Omega^{'}\subseteq{\Omega}$,    $\Xi=\{\Sigma,{T},\Omega,\varphi\}$ is stabilizable  from $S_0$ at time $t_0\in{T}$ if $\Xi^{'}=\{\Sigma,\Theta,\Omega^{'},\varphi\}$ is $S_1$-stabilizable from $S_0$ at time  $t_0\in{T}$.

($P_3$) A dynamical system  $\Xi=\{\Sigma,{T},\Omega,\varphi\}$ is controllable at time  $t_0\in{T}$ if and only if it  is $\Sigma$-stabilizable from $\Sigma$ at time  $t_0\in{T}$.

($P_4$) Let $\Sigma_1$ is a subspace of $\Sigma$,  a dynamical system  $\Xi=\{\Sigma,{T},\Omega,\varphi\}$ is controllable on $\Sigma_1$  at time  $t_0\in{T}$ if and only if it  is $\Sigma_1$-stabilizable from $\Sigma_1$ at time  $t_0\in{T}$.

($P_5$) The state $x_1$ of  dynamical system $\Xi=\{\Sigma,{T},\Omega,\varphi\}$ is stabilizable  from $S_0$ at time $t_0\in{T}$ if it  is $\{x_1\}$-stabilizable from $S_0$ at time  $t_0\in{T}$.

 The proof of the aforementioned properties are quite straightforward, but the properties are of major importance: set stabilizability  is regarded as the generalization of both controllability and stabilizability.

  Because we need to steer systems within the specified finite time in many applications, we further propose the notation  of set stabilizability within the specified time span $T_s$ as follows.

\textbf{Definition 5}. Given  $T_s>0$, the dynamical system  $\Xi=\{\Sigma,{T},\Omega,\varphi\}$ is $S_1$-stabilizable within the specified time $t_s$ from $S_0$ at time  $t_0\in{T}$, if  for any  initial state $x_0\in{S_0}$ and another arbitrary target state $x_1\in{S_1}$, there exist $t_1\in{T}$ with ${t_1-t_0}\leq{T_s}$ and  admissible control $u\in\Omega$ such that $x_1=\varphi_{u}(t_1;t_0,x_0)$ and $\varphi_{u}(t;t_0,x_0)\in{S_1}$ when $t_1\leq{t}\in{T}$.

\subsection{Equilibrium points of Schr$\ddot{o}$dinger Equation}

In this paper, we will study  a special subclass of dynamical systems, those which are quantum closed systems, to illustrate why set-stabilizable is useful.

Before investigating stabilizability of quantum closed systems, we will have to discuss how to define an equilibrium point  of Schr$\ddot{o}$dinger Equation
\begin{equation}
\label{S}
i\hbar\frac{d}{dt}|\psi(t)\rangle=H|\psi(t)\rangle
\end{equation}
where $|\psi(t)\rangle$ is a pure state in Hilbert space. For the purpose of simplicity, we set $\hbar=1$ in the whole paper.

From the mathematical point of view, it seems that  zero vector is  the sole  equilibrium point of Eq. (\ref{S}). Unfortunately, zero vector is nonsense from the  viewpoint of physics. We need to give some further investigation on this issue.  To obtain some intuitive pictures about the solution of Eq. (\ref{S}), let us consider a two-level quantum system given by
\begin{equation}
\label{qubit}
i\frac{d}{dt}|\psi(t)\rangle=\sigma_z|\psi(t)\rangle
\end{equation}
where $\sigma_z=|0\rangle\langle0|-|1\rangle\langle1|$. From the physical point of view, both $|0\rangle$ and $|1\rangle$ are the equilibrium points of Eq. (\ref{qubit}).  From the mathematical point of view, the solution $|\psi(t)\rangle$ of  Eq. (\ref{qubit}) with the initial state $|\psi(0)\rangle=|0\rangle$ (or $|\psi(0)\rangle=|1\rangle$) satisfies $|\psi(t)\rangle\langle\psi(t)|=|0\rangle\langle0|$ (or $|\psi(t)\rangle\langle\psi(t)|=|1\rangle\langle1|$). In other words, $|\psi(t)\rangle\langle\psi(t)|\equiv|\psi(0)\rangle\langle\psi(0)|$ if $|\psi(0)\rangle=|i\rangle$ with $i=0,1$.  It is also revealed  that $[\sigma_z,|i\rangle\langle{i}|]=0$ for $i=0,1$, where $[\ ,\ ]$ is Lie bracket which is specified by $[A,B]=AB-BA$.

Based on aforementioned observations, we have the following definition:

\textbf{Definition 6:} $|\psi_s\rangle$ is called an equilibrium point  of Eq. (\ref{S}) if $[H,|\psi_s\rangle\langle\psi_s|]=0$.

\textbf{Remark:}  Denote $\rho(t)=|\psi(t)\rangle\langle\psi(t)|$, then Eq. (\ref{S}) can be written as $\frac{d}{dt}\rho(t)=[H,\rho(t)]$. This implies that $\rho(t)=\rho_s=|\psi_s\rangle\langle\psi_s|$ is the static solution of $\frac{d}{dt}\rho(t)=[H,\rho(t)]$ if $|\psi_s\rangle$ is called  the equilibrium point of Eq. (\ref{S}).

\textbf{Proposition 2:} $|\psi_s\rangle$ is  an equilibrium point  of Eq. (\ref{S}) if and only if $|\psi_s\rangle$ is an eigenvector of  Hamiltonian $H$.

\subsection{Set stabilizability for quantum closed systems}

Consider a controlled finite-dimensional quantum system without output
\begin{equation}
\label{1}
i\frac{d}{dt}|\psi(t)\rangle=[H_0+H_{c}(t)]|\psi(t)\rangle
\end{equation}
where $|\psi(t)\rangle$ is a pure state in Hilbert space, and $H_0$ and $H_c$ are system Hamiltonian and controlled Hamiltonian,  respectively.

For the controlled quantum system governed by (\ref{1}),  set stabilizability can be specially defined as follows:

\textbf{Definition 4a}:  For  an  arbitrary initial state $|\psi_0\rangle\in{S_0}$ and another arbitrary target state  $|\psi_f\rangle\in{S_1}$, the quantum system  Eq. (\ref{1}) is $S_1$-stabilizable   from $S_0$ at time $t_0$ if  there exist a finite time $t_1\geq{t_0}$ and permissible control $H_{c}(t)$ such that  the system can be transferred from $|\psi(t_0)\rangle=|\psi_0\rangle\in{S_0}$ to $|\psi(t_1)\rangle=|\psi_f\rangle$ and $|\psi(t)\rangle\in{S_1}$  when  $t\geq{t_1}$.

 To overcome decoherence, we need to steer quantum systems within the specified time span $T_s$. In this situation, we can present the notion of set stabilizability within the specified time span $T_s$ by modifying the Definition 4a.

\textbf{Definition 5a:}   Given  $T_s>0$, $\forall$ $|\psi_0\rangle\in{S_0}$ and  $|\psi_f\rangle\in{S_1}$, the quantum system  Eq. (\ref{1}) is $S_1$-stabilized within the specified time span $T_s$  from $S_0$ at time $t_0$ if  there exist a finite time $t_1$ with ${t_1-t_0}\leq{T_s}$ and permissible control $H_{c}(t)$ such that  the system can be transferred from $|\psi(t_0)\rangle=|\psi_0\rangle\in{S_0}$ to $|\psi(t_1)\rangle=|\psi_f\rangle$ and $|\psi(t)\rangle\in{S_1}$  when  $t\geq{t_1}$.

\section{Set stabilizability for two-level quantum systems}
In this section, we will focus on set stabilizability of two-level quantum systems.

\subsection{Model description and notation}
Let $|0\rangle$ and $|1\rangle$ be a basis of Hilbert space of  two-level  quantum systems. A two-level quantum system is governed by
\begin{equation}
\label{2}
\frac{d}{dt}|\psi(t)\rangle={i}[\omega_0S_{z}+u_{x}(t)S_{x}+u_{y}(t)S_{y}]|\psi(t)\rangle
\end{equation}
where $S_{z}=\frac{1}{2}\sigma_{z}=\frac{1}{2}(|0\rangle\langle0|-|1\rangle\langle1|)$,
$S_{x}=\frac{1}{2}\sigma_{x}=\frac{1}{2}(|1\rangle\langle0|+|0\rangle\langle1|)$,
$S_{y}=\frac{1}{2}\sigma_{y}=\frac{i}{2}(|1\rangle\langle0|-|0\rangle\langle1|)$,  $u_x(t)$ and $u_{y}(t)$ are adjustable scale functions.
and  $|u_{x}(t)|\leq{L_x}$ and $|u_{y}(t)|\leq{L_y}$. For simplicity, we assume that  $L_x=L_y=g_0$ in this paper.

The state space is denoted as $\Sigma=span\{|0\rangle,|1\rangle\}$ and can be further parameterized  as:
\begin{equation}
\label{2-1}
\Sigma=\{|\psi\rangle=\cos\frac{\theta}{2}|0\rangle+e^{i\phi}\sin\frac{\theta}{2}|1\rangle: \theta\in[0,\pi],\phi\in[0,2\pi)\}
\end{equation}
where $\theta$ and $\phi$ are Bloch parameters.

For the further discussions, we introduce the following notations:
\begin{equation}
\label{2-1circle}
C_{\theta_f}=\{|\psi\rangle=\cos\frac{\theta_f}{2}|0\rangle+e^{i\phi}\sin\frac{\theta_f}{2}|1\rangle:\phi\in[0,2\pi)\}
\end{equation}
and
\begin{equation}
\label{2-1point}
P_{\theta_f,\phi_f}=\{|\psi_{f}\rangle=\cos\frac{\theta_f}{2}|0\rangle+e^{i\phi_f}\sin\frac{\theta_f}{2}|1\rangle\}
\end{equation}
 where $C_{\theta_f}$ represents a circle on a Bloch sphere, $P_{\theta_f,\phi_f}$ is regarded as a point on the circle $\Sigma_{\theta_f}$,  $\theta_f$ is a fixed value in $[0,\pi]$ and $\phi_f$ is a fixed value in $[0,2\pi)$.

\begin{figure}[ht]
\centering \subfigure[$C_{\theta_f}$: a circle on a Bloch sphere] {
\label{Fig.4:a} %% label for first subfigure
 \scalebox{0.3}{\includegraphics{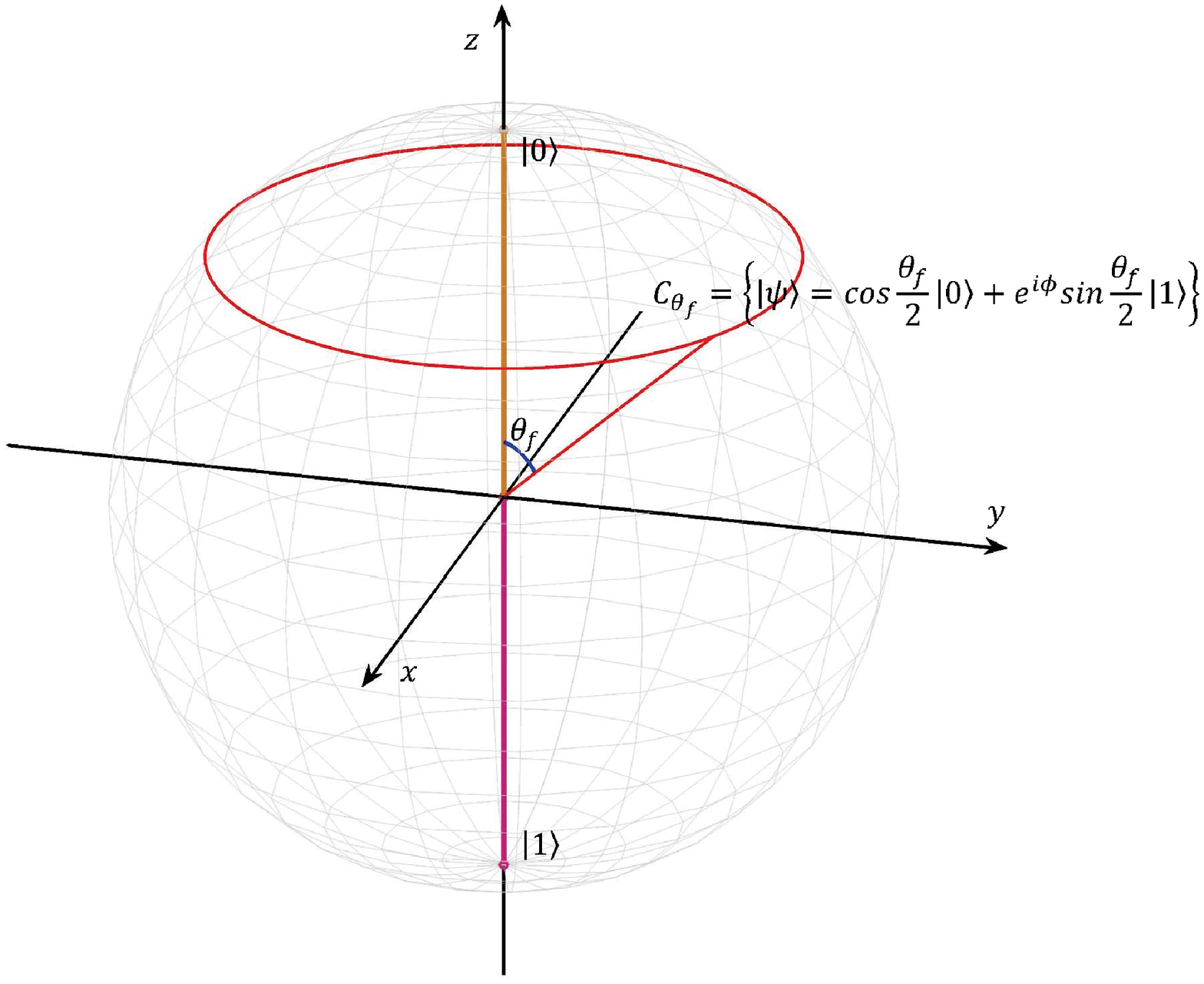}}}%
\subfigure[$P_{\theta_f,\phi_f}$:a point on a Bloch sphere] {
\label{Fig.4:b} %% label for second subfigure
 \scalebox{0.4}{\includegraphics{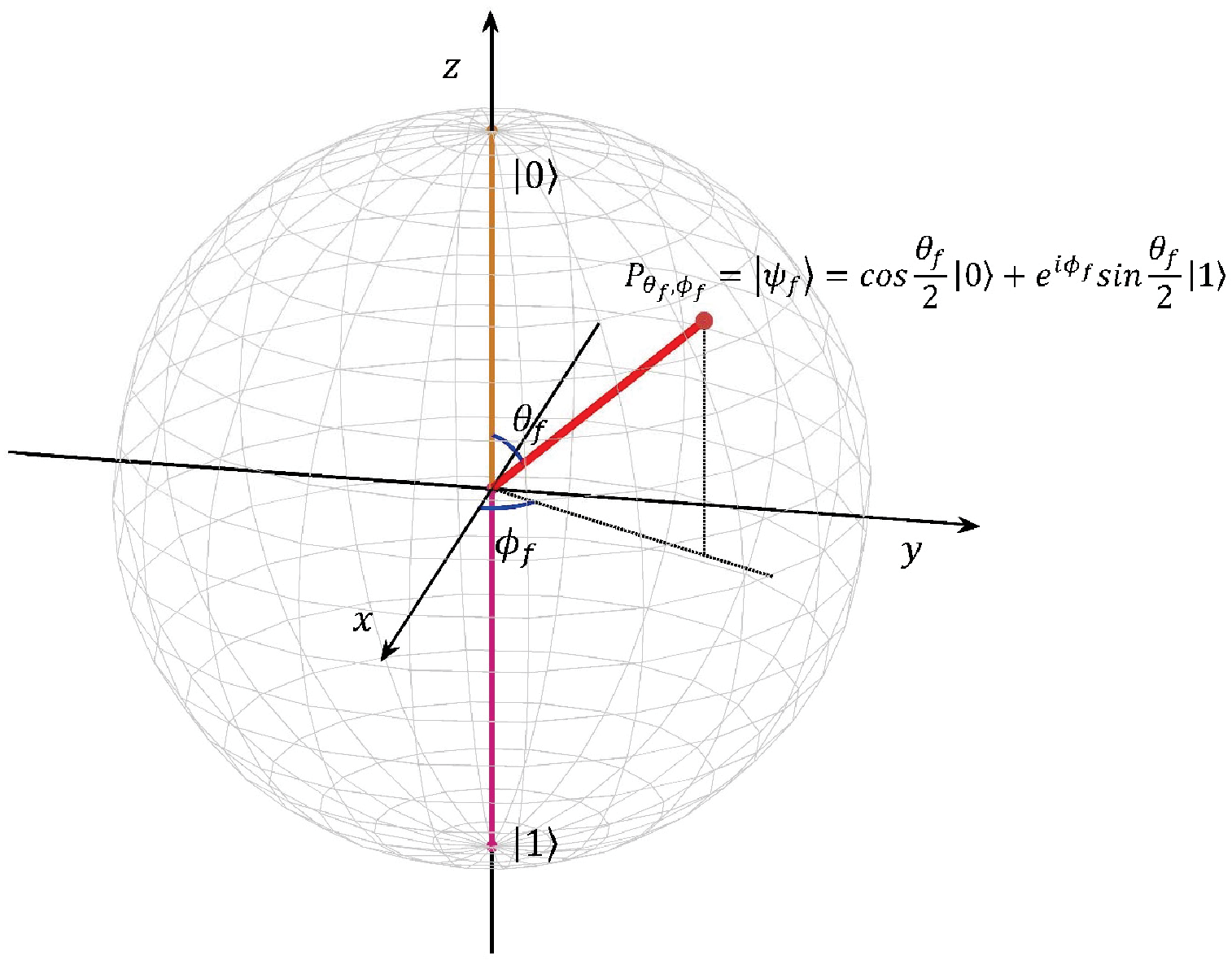}}
}%
\caption{\label{fig4} Geometric representation of $C_{\theta_f}$ and $P_{\theta_f,\phi_f}$ } %% label for entire figure
\end{figure}

 To obtain some intuitive pictures of $C_{\theta_f}$ and $P_{\theta_f,\phi_f}$, we plot them in Fig \ref{fig4}.

 Two kind of permission control sets are considered: bounded control with bound $g_0$
$\Omega_{B}(g_0)=\{u_x(t)S_x +u_y(t)S_y:|u_x(t)|\leq{g_0},|u_y(t)|\leq{g_0}\}$ and bounded time-continuous controls with bound $g_0$  $\Omega_{BC}({g_0})=\{u_x(t)S_x +u_y(t)S_y:u_x(t),u_y(t)\in{C^{0}},|u_x(t)|\leq{g_0},|u_y(t)|\leq{g_0}\}$
where $C^{0}$ is the space of time-continuous functions.

\subsection{Stabilizability and circle-set stabilizability for two-level quantum systems}
In this subsection, we will first establish the necessary and sufficient conditions for that the given target state is stabilizable by bounded controls in $\Omega_{B}(g_0)$.

\textbf{{Theorem 1}}: Given $\theta_f\in[0,\pi]$ and $\phi_f\in[0,2\pi)$, $\forall{t_0}>0$, the controlled qubit system  Eq. (\ref{2})  is $P_{\theta_f,\phi_f}$-stabilizable from $\Sigma$ at $t_0$ by $\Omega_{B}(g_0)$ if and only if
\begin{equation}
\label{4-0}
\omega_0\cdot|\tan\theta_f|\cdot\max\{|\sin\phi_f|,|\cos\phi_f|\}\leq{g_0}
\end{equation}

\textbf{Proof:}  See in Appendix.

Subsequently, we will further give a theorem for circle-set stabilizability of two-level quantum systems.

\textbf{{Theorem 2}}: For $\forall\theta_f\in{[0,\pi]}$, $\forall{g_0>0}$ and $\forall{t_0>0}$, the controlled quantum system  Eq. (\ref{2})  is $C_{\theta_f}$-stabilizable from $\Sigma$ by $\Omega_{BC}(g_0)$.

\textbf{\emph{{Proof}}}: See in Appendix.

\textbf{Remark 2:}  It should be underlined that $|\psi(t)\rangle=\cos{\frac{\theta_{f}}{2}}|0\rangle+e^{i(\phi_f-2k\pi)}\sin{\frac{\theta_{f}}{2}}|1\rangle=|\psi_f\rangle$ when $t=t_f+\frac{2k\pi}{\omega_0}$ where $k\in{Z^{+}}$. This means that for   all $|\psi_f\rangle\in{\Sigma}$, it can be dynamically stored by permissible control in $\Omega_{BC}(g_0)$. We immediately conclude that  the controlled two-level quantum  system  Eq. (\ref{2})  is $C_{\theta_f}$-stabilizable from $\Sigma$ by $\Omega_{B}(g_0)$  for all $\theta_f\in{[0,\pi]}$ since $\Omega_{BC}(g_0)$ is the subset of $\Omega_{B}(g_0)$.

\subsection{Circle-set stabilizability within the specified time for two-level quantum systems}

In this subsection, we will establish a theorem to explore whether or not the system (\ref{2}) is  $C_{\theta_s}$-stabilized  from $\Sigma$ within the specified time $T_s$ by  admissible controls in $\Omega_{BC}({g_0})$.

\textbf{{Theorem 3}}:   For any $\theta_f\in{[0,\pi]}$ and any ${t_0}>0$, the controlled two-level system  Eq. (\ref{2})  is $C_{\theta_f}$-stabilized within the specified time $T_s$ from $\Sigma$  at $t_0$ by $\Omega_{BC}(g_0)$ if
\begin{equation}
\label{theorem2}
{\frac{4\pi}{g_0}+\frac{8\pi}{\omega_0}}\leq{{T_s}}.
\end{equation}

%\textbf{Remark 3}: If the permitted controls are chosen from $\Omega_{BC}(+\infty)$, we have
%\begin{equation}
%\label{R2-1}
%t_f-t_0\leq\frac{4\pi}{\omega_0}.
%\end{equation}
%for any pair of initial and target states. For $\forall$ $\theta_f\in{[0,\pi]}$ and $\forall{t_0}>0$. the controlled qubit system  Eq. (\ref{2})  is $C_{\theta_f}$-stabilizable within the specified time $T_s$ from $\Sigma$  at $t_0$ by $\Omega_{BC}(+\infty)$ if
%\begin{equation}
%\label{Remark2}
%{\frac{4\pi}{\omega_0}}\leq{T_s}.
%\end{equation}

Furthermore,  Eq. (\ref{theorem2}) can be  improved if the permissible controls are chosen from $\Omega_{B}(g_0)$ instead of $\Omega_{BC}(g_0)$.

\textbf{{Theorem 4}}:  Let $\theta_f\in[0,\pi]$ and $\phi_f\in[0,2\pi)$, $\forall{t_0}>0$, the controlled two-level system  Eq. (\ref{2})  is $C_{\theta_f}$-stabilizable within the specified time $T_s$ from $\Sigma$ at $t_0$ by $\Omega_{B}(g_0)$ if
\begin{equation}
 \label{theorem3-1}
 {\frac{\pi}{g_0}+\frac{8\pi}{\omega_0}}\leq{T_s}.
\end{equation}
or
\begin{equation}
 \label{theorem4-1}
 {\frac{4\pi}{g_0}+\frac{6\pi}{\omega_0}}\leq{T_s}.
\end{equation}

\subsection{Circle-set stabilization of two-level quantum systems with multiple constrains}

In this subsection, we  first exploit the sufficient conditions for set-stabilizability with both time constrain $T_s$ and energy constrain $E_s$.

 For the given $\theta_0$, $\phi_0$, $\theta_f$ and $\phi_f$, we will investigate whether or not  there exist permissible  controls  such that  the controlled qubit system  Eq. (\ref{2}) is $C_{\theta_f}$-stabilizable with the following constrained conditions
\begin{equation}
 \label{Time}
t_f-t_0\leq{T_s}
\end{equation}
and
\begin{equation}
 \label{Energy}
\int^{t_f}_{t_0}[u^{2}_x(t)+u^{2}_y(t)]dt\leq{E_s},
\end{equation}

When the unbounded control are permitted, we have the following theorem:

\textbf{Theorem 5}:  Given $T_s$ and $E_s$, for $\forall$ $\theta_0$, $\phi_0$, $\theta_f$ and $\phi_f$,  there exist unbounded  controls  such that  the controlled qubit system  Eq. (\ref{2}) is $C_{\theta_f}$-stabilizable with time-energy constrains Eqs. (\ref{Time}) and  (\ref{Energy})
 if $T_s\geq\frac{7\pi}{\omega_0}$ and $E_s\geq\omega_0\cdot\pi$.

\textbf{Proof}:  If $T_s\geq\frac{7\pi}{\omega_0}$ and $E_s\geq\omega_0\cdot\pi$, then there exists at least $k=2$ such that
\begin{equation}
 \label{E1}
\frac{\omega_0\pi\sin^{2}\frac{\theta_0+\theta_f}{2}}{2E_s}+\frac{\phi_f-\phi_0}{2\pi}-\frac{\cos\frac{\theta_0+\theta_f}{2}}{2}\leq{k}
\end{equation}
and
\begin{equation}
 \label{T1}
{k}\leq\frac{{T_s}{\omega_0}+\phi_f-\phi_0-\pi\cos\frac{\theta_0+\theta_f}{2}}{2\pi}.
\end{equation}
and
\begin{equation}
 \label{plus}
2k\pi-\phi_f+\phi_0+\pi\cos\frac{\theta_0+\theta_f}{2}>0
\end{equation}
 hold for $\forall$ $\theta_0$, $\phi_0$, $\theta_f$ and $\phi_f$.

Therefore, we  establish the sufficient conditions for $C_{\theta_f}$-stabilizable with time-energy constrains.

\section{Stabilizability and set stabilizability of quantum systems}
The necessary and sufficient conditions  can be established for that the state $|\psi_f\rangle$ of quantum system  Eq. (\ref{1}) is stabilizable  from $S_0$.

\textbf{Theorem 6}:  The state $|\psi_f\rangle$ of quantum system  Eq. (\ref{1}) is stabilizable from $S_0$ if and only if two following conditions are satisfied:

(R1) $|\psi_f\rangle$ is reachable from any state in $S_0$ by  a coherent control $H_c(t)$;

 (E2) there exists a static control Hamiltonian $H_c$ such that $|\psi_s\rangle$ is an eigenvector of  Hamiltonian $H_0+H_c$.

\begin{figure}[ht]
\centering \subfigure[$\frac{g_0}{\omega_0}=0.1$] {
\label{Fig.3:a} %% label for first subfigure
 \scalebox{0.3}{\includegraphics{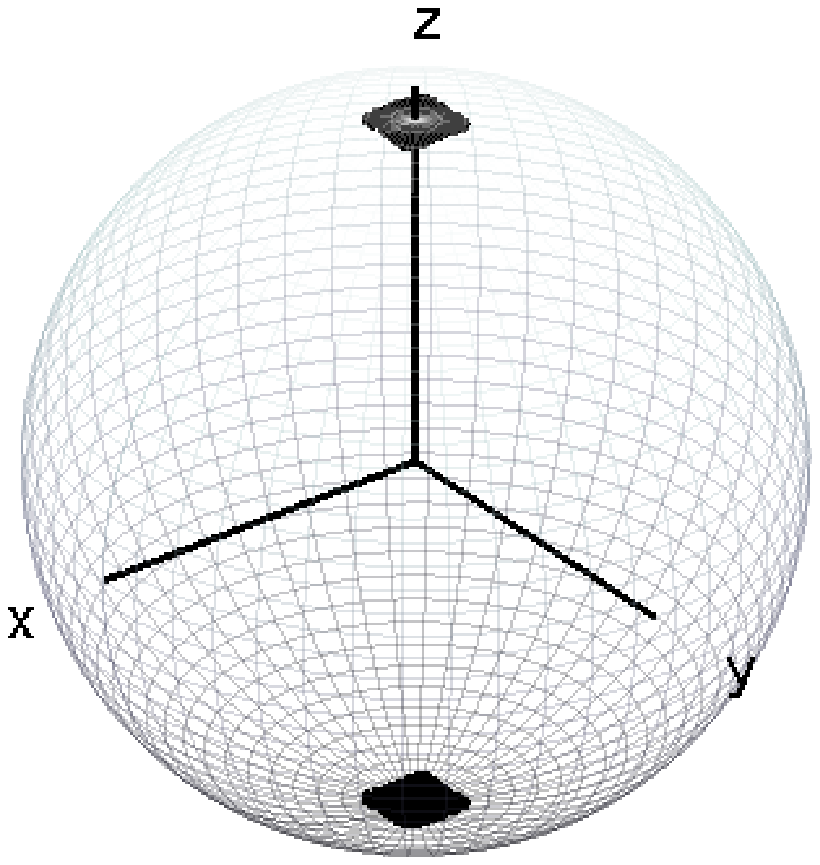}}}%
\subfigure[$\frac{g_0}{\omega_0}=0.2$] {
\label{Fig.3:b} %% label for second subfigure
 \scalebox{0.3}{\includegraphics{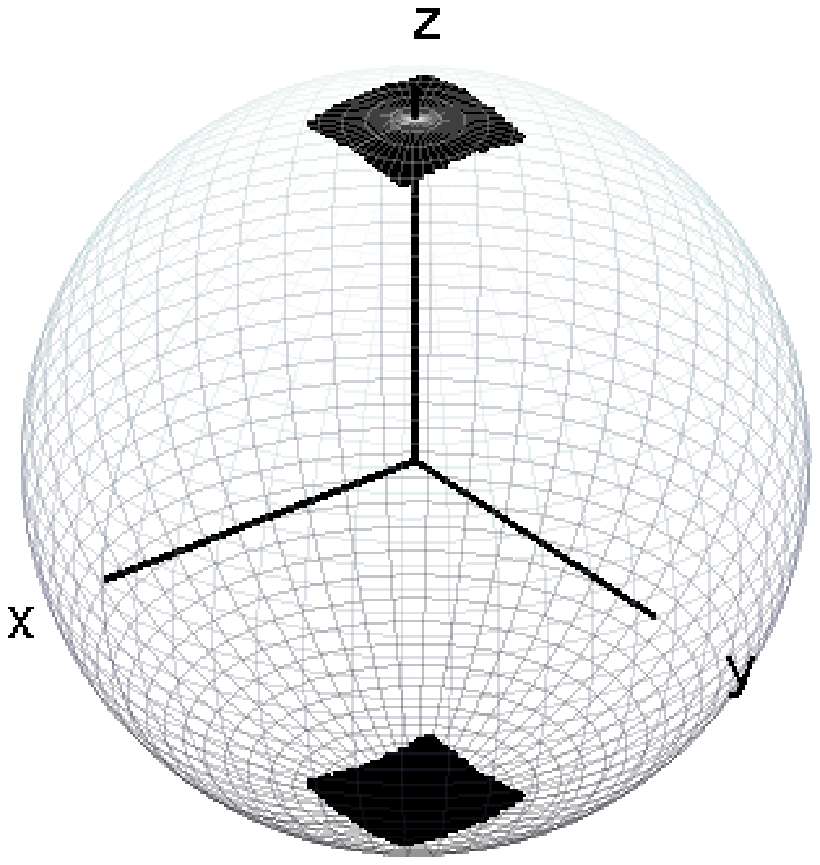}}
}%
\\
\centering \subfigure[$\frac{g_0}{\omega_0}=0.5$] {
 \label{Fig.3:c} %% label for third subfigure
\scalebox{0.3}{\includegraphics{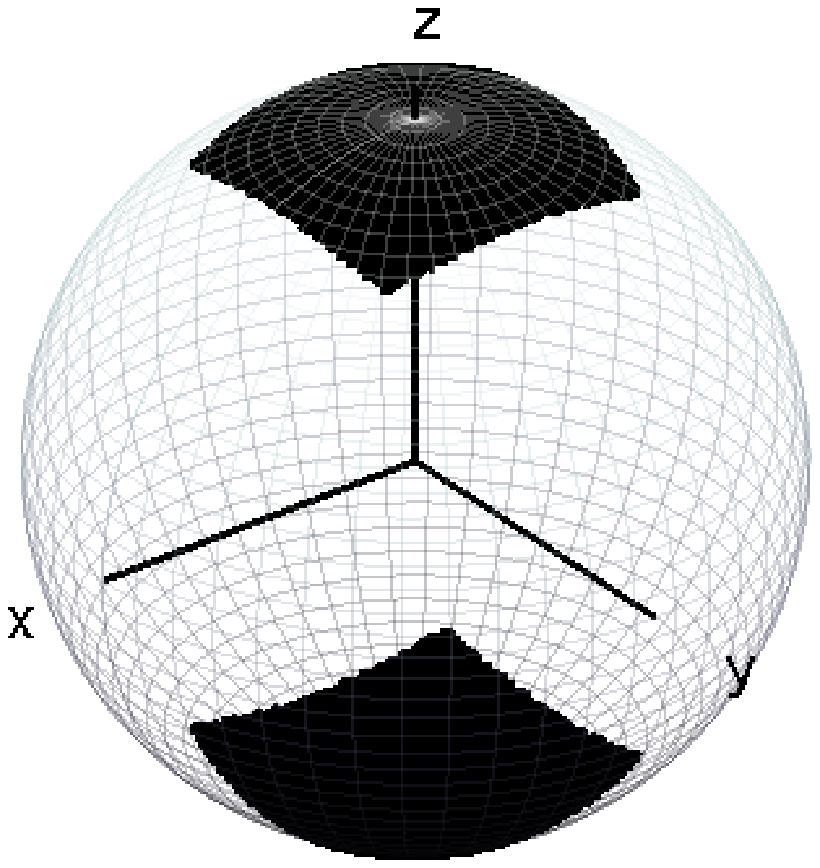}}}
\subfigure[$\frac{g_0}{\omega_0}=1$] {
 \label{Fig.3:d} %% label for fourth subfigure
\scalebox{0.3}{\includegraphics{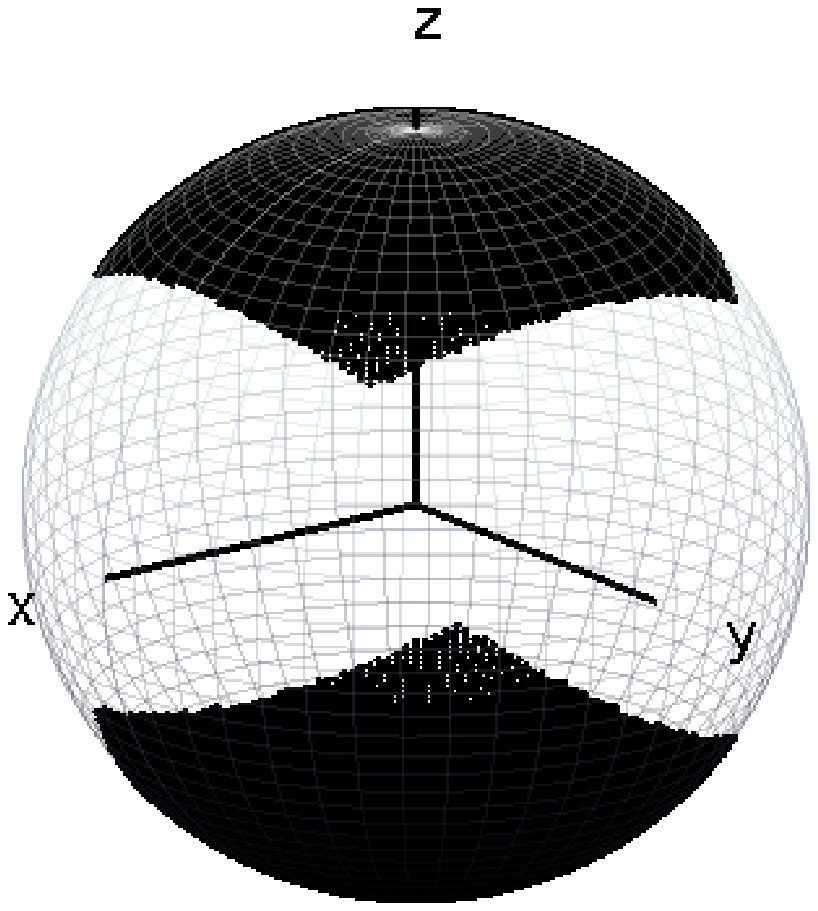}}}%
\caption{\label{fig3} Stabilizable state regions with different ratio $\frac{g_0}{\omega_0}$} %% label for entire figure
\end{figure}

\emph{Remark:} Unfortunately, the  condition (E2) is very difficult to be satisfied. For two-level quantum systems Eq. (\ref{1}),  the  condition (E2) is reduced to Eq. (\ref{4-0}). To obtain some intuitive pictures of how   Eq. (\ref{4-0}) is quite strict, we write Eq. (\ref{4-0})   as
\begin{equation}
\label{4-0m}
|\tan\theta_f|\cdot\max\{|\sin\phi_f|,|\cos\phi_f|\}\leq\frac{g_0}{\omega_0}.
\end{equation}
and  plot the range of stabilizable states with ratio $\frac{g_0}{\omega_0}=0.1,0.2,0.5,1$ in Fig \ref{fig3}.

 Because $\frac{g_0}{\omega_0}\leq10^{-3}\ll{1}$ holds for nuclear spin systems\cite{R1}, we immediately realize from Fig \ref{fig3} that   Eq. (\ref{4-0}) is very strict in some experimental quantum systems.

  This observation about stabilizability is in remarkable contrast with that about set-stabilizability in Theorem 2:   the controlled two-level quantum system  Eq. (\ref{2})  is $C_{\theta_f}$-stabilizable from $\Sigma$ at $t_0$ by $\Omega_{BC}(g_0)$ for $\forall\theta_f\in{[0,\pi]}$, $\forall{g_0>0}$ and $\forall{t_0>0}$.

  For  two-level quantum system
\begin{equation}
\label{2x}
\frac{d}{dt}|\psi(t)\rangle={i}[\omega_0S_{z}+u_{x}(t)S_{x}]|\psi(t)\rangle
\end{equation}
 the necessary condition for   (E2) is that $|\psi_f\rangle=\cos\frac{\theta_f}{2}|0\rangle+e^{i\phi_f}\sin\frac{\theta_f}{2}|1\rangle$ with $\phi_f=0,\pi$, this is also in remarkable contrast with the observation that the controlled two-level quantum system  Eq. (\ref{2x})  is $C_{\theta_f}$-stabilizable from $\Sigma$ at $t_0$ by bounded control $|u_{x}|\leq{g_0}$ for $\forall\theta_f\in{[0,\pi]}$, $\forall{g_0>0}$ and $\forall{t_0>0}$.

 For  two-level quantum system
\begin{equation}
\label{2y}
\frac{d}{dt}|\psi(t)\rangle={i}[\omega_0S_{z}+u_{y}(t)S_{y}]|\psi(t)\rangle
\end{equation}
the necessary condition for   (E2) is that $|\psi_f\rangle=\cos\frac{\theta_f}{2}|0\rangle+e^{i\phi_f}\sin\frac{\theta_f}{2}|1\rangle$ with $\phi_f=\frac{\pi}{2}, \frac{3\pi}{2}$. this is also in remarkable contrast with the observation that the controlled two-level quantum system  Eq. (\ref{2y})  is $C_{\theta_f}$-stabilizable from $\Sigma$ at $t_0$ by bounded control $|u_{y}|\leq{g_0}$ for $\forall\theta_f\in{[0,\pi]}$, $\forall{g_0>0}$ and $\forall{t_0>0}$.

From the aforemention discussions, we realize that  the necessary conditions are too strict for stabilizability of quantum systems but some kinds of set stabilizability notions are available.

We further present another example to illustrate that sufficient conditions are easier to be fulfilled for set stabilizability.

 Consider a controlled two-qubit system governed by the equation
\begin{equation}
\label{5-1}
i\frac{d}{dt}|\psi(t)\rangle=[H_0+H_c(t)]|\psi(t)\rangle
\end{equation}
where $H_0=-\omega_0\sigma^{1}_z+\omega_0\sigma^{2}_z$ is system Hamiltonian and $H_c(t)=\sum_{i,j=x,y}{u_{ij}(t)\sigma^{1}_{i}\otimes\sigma^{2}_{j}}$ are  controlled Hamiltonian,  respectively.

Let  $\Sigma_{s}=span\{|0_{1}1_2\rangle,|1_{1}0_2\rangle\}$ be a subspace for two-qubit system, we have
\begin{equation}
\label{5-2s}
\Sigma_{s}=\{\cos\frac{\theta}{2}|0_{1}1_2\rangle+e^{i\phi}\sin\frac{\theta}{2}|1_{1}0_2\rangle:\theta\in[0,\pi],\phi\in[0,2\pi)\}
\end{equation}
Denote a maximal  entangled state  subset as
\begin{equation}
\label{5-2e}
E_{M}=\{\frac{\sqrt{2}}{2}(|0_{1}1_2\rangle+e^{i\phi}|1_{1}0_2\rangle):\phi\in[0,2\pi)\}
\end{equation}

Introducing $|0^{L}\rangle=|0_{1}1_2\rangle$ and  $|1^{L}\rangle=|1_{1}0_2\rangle$, we have
\begin{equation}
\label{5-2Lz}
\sigma^{L}_{z}=|0^{L}\rangle\langle0^{L}|-|1^{L}\rangle\langle1^{L}|=\frac{1}{2}(\sigma^{1}_z\otimes{I^{2}}-{I^{1}}\otimes\sigma^{2}_z)
\end{equation}
and
\begin{equation}
\label{5-2Lx}
\sigma^{L}_{x}=|1^{L}\rangle\langle0^{L}|+|0^{L}\rangle\langle1^{L}|=\frac{1}{2}(\sigma^{1}_{x}\otimes\sigma^{2}_{x}+\sigma^{1}_{y}\otimes\sigma^{2}_{y})
\end{equation}
and
\begin{equation}
\label{5-2Ly}
\sigma^{L}_{y}={i}(|1^{L}\rangle\langle0^{L}|-|0^{L}\rangle\langle1^{L}|)=\frac{1}{2}(\sigma^{1}_{y}\otimes\sigma^{2}_{x}+\sigma^{1}_{x}\otimes\sigma^{2}_{y})
\end{equation}

From the geometric point of view, the subspace $\Sigma_{s}$ can be regarded as a Bloch sphere of the encoded qubit
\begin{equation}
\label{5-2Ls}
\Sigma_{s}=\{\cos\frac{\theta}{2}|0^{L}\rangle+e^{i\phi}\sin\frac{\theta}{2}|1^{L}\rangle:\theta\in[0,\pi],\phi\in[0,2\pi)\}
\end{equation}
and the maximal  entangled state  subset \begin{equation}
\label{5-2EM}
E_M=\{\cos\frac{\pi}{4}|0^{L}\rangle+e^{i\phi}\sin\frac{\pi}{4}|1^{L}\rangle:\phi\in[0,2\pi)\}
\end{equation}
 can be treated as a circle on the Bloch sphere.

Let $S^{L}_{i}=\frac{1}{2}\sigma^{L}_{i}$ with $i=x,y,z$,  $u_{xx}(t)=u_{yy}(t)=u^{L}_x(t)$ and  $u_{xy}(t)=-u_{yx}(t)=u^{L}_y(t)$,   then Eq. (\ref{5-1}) can be rewritten as
\begin{equation}
\label{5-3}
i\frac{d}{dt}|\psi^{L}(t)\rangle=4[-\omega_0S^{L}_{z}+u^{L}_{x}(t)S^{L}_{x}+u^{L}_{y}(t)S^{L}_{y}]|\psi^{L}(t)\rangle
\end{equation}

Let $\Omega^{L}_{BC}(g_0)=\{\sum_{i,j=x,y}{u_{ij}(t)\sigma^{1}_{i}\otimes\sigma^{2}_{j}}:u_{ij}(t)\in{C^{0}}, |u_{ij}|\leq{g_0}\}$ and $\Omega^{L}_{B}(g_0)=\{\sum_{i,j=x,y}{u_{ij}(t)\sigma^{1}_{i}\otimes\sigma^{2}_{j}}: |u_{ij}|\leq{g_0}\}$ for $i,j=x,y$, we have the following Corollaries:

\textbf{Corollary 1}:   For $\forall{g_0>0}$ and $\forall{t_0>0}$, the controlled qubit system  Eq. (\ref{5-1})  is $E_{M}$-stabilizable from $\Sigma_s$ at time $t_0$ by $\Omega^{L}_{BC}(g_0)$.

\textbf{Corollary 2}:  For $\forall{t_0>0}$, the controlled qubit system  Eq. (\ref{5-1})  is $E_{M}$-stabilizable within $T_s$ from $\Sigma_s$ at time $t_0$ by $\Omega^{L}_{BC}(g_0)$ if  ${\frac{\pi}{g_0}+\frac{2\pi}{\omega_0}}\leq{T_s}$.

\textbf{Corollary 3}: For $\forall{t_0>0}$, the controlled qubit system  Eq. (\ref{5-1})  is $E_{M}$-stabilizable within $T_s$ from $\Sigma_s$ at time $t_0$ by $\Omega^{L}_{B}(g_0)$ if  ${\frac{\pi}{4g_0}+\frac{2\pi}{\omega_0}}\leq{T_s}$ or  ${\frac{\pi}{g_0}+\frac{3\pi}{2\omega_0}}\leq{T_s}$.

The aforementioned results suggest that set-stabilizability  can  be used for studying entangled state generation problem.

\section{Conclusions}
In summery, we explored set-stabilizability by constrained open-loop controls in this research. Both controllability and stabilizability can be regarded as the special case of set-stabilizability.  The necessary and sufficient conditions are also established for stabilizability of quantum closed systems, and it is further revealed  that the necessary conditions  are too strict for stabilization of some concrete quantum systems like nuclear spin systems. This further justifies  the introduction of  the  set-stabilizability notion.

 We also  clarify how to define an equilibrium point of Schr$\ddot{o}$dinger Equation from the physical point of view.  Strictly speaking, stabilizability problems for quantum systems should be discussed in terms of density operators and master equations. It is  exemplified  that set-stabilizability  can  be used for investigating quantum information processing problems including quantum information storage and entangled state generation.

 In our opinion, it should be further investigated that what kind of set-stabilizability is achieved for quantum open systems.

\begin{ack}                               % Place acknowledgements
 Partially supported by the National Nature Science
Foundation of China under Grant Nos. 61273202 and 61134008.   % here.
\end{ack}

\section{Appendix}
 \subsection{Proof of Theorem 1}
(1) First, we need to prove that if Eq. (\ref{4-0}) holds,  there always exist  bounded  controls in $\Omega_{B}(g_0)$
 to transit the qubit system from an arbitrary initial state
$|\psi(t_0)\rangle=|\psi_0\rangle=\cos\frac{\theta_0}{2}|0\rangle+e^{i\phi_0}\sin\frac{\theta_0}{2}|1\rangle\in\Sigma$
to another arbitrary target state
$|\psi(t_{f})\rangle=|\psi_f\rangle=\cos\frac{\theta_f}{2}|0\rangle+e^{i\phi_f}\sin\frac{\theta_f}{2}|1\rangle$ with $t_0<t_{f}<+\infty$,
and  $|\psi(t)\rangle\langle\psi(t)|=|\psi_f\rangle\langle\psi_f|$ with $t\geq{t_f}$.

Choose the permissible controls  as follows:
\begin{equation}
\label{T5-01}
u_{x}(t)=\left\{\begin{array}{ll}
g\cos[\omega_{rf}(t-t_0)+\varphi_1]& t\in[t_0,t_{f})\\
\omega_0\cdot\tan\theta_f\cdot\cos\phi_f& t\in[t_f,+\infty)
\end{array}\right.
\end{equation}
and
\begin{equation}
\label{T5-02}
u_{y}(t)=
\left\{\begin{array}{ll}
g\sin[\omega_{rf}(t-t_0)+\varphi_1]& t\in[t_0,t_{f})\\
\omega_0\cdot\tan\theta_f\cdot\sin\phi_f&t\in[t_f,+\infty)
\end{array}\right.
\end{equation}
where $g\in[0.g_0]$ and $\omega_{rf}\in{R}$, and the designed parameters are given by
\begin{equation}
\label{T3-03}
\varphi_{1}=\phi_0,
\end{equation}
\begin{equation}
\label{T3-04}
g=\frac{\omega_0\pi\sin\frac{\theta_0+\theta_f}{2}}{\phi^{fap}_{k}},
\end{equation}
\begin{equation}
\label{T3-05}
\omega_{rf}=\frac{-(2k\pi-\phi_{f}+\phi_0)\omega_0}{\phi^{fap}_{k}}
\end{equation}
and
\begin{equation}
\label{T3-06}t_f=\frac{\phi^{fap}_{k}}{\omega_0}+t_0
\end{equation}
where
\begin{equation}
\label{T3-07}
\phi^{fap}_{k}=2k_{fap}\pi-\phi_{f}+\phi_{0}+\pi{\cos\frac{\theta_{0}+\theta_{f}}{2}}
\end{equation}
and $k_{fap}$ is such an  integer that
\begin{equation}
\label{T3-08}
k_{fap}\geq\frac{\phi_f-\phi_0-\pi\cos\frac{\theta_0+\theta_f}{2}}{2\pi}+\frac{\omega_0\sin\frac{\theta_0+\theta_f}{2}}{2g_0}
\end{equation}

i) It is demonstrated by some calculations that  $|\psi(t_{f})\rangle=\cos{\frac{\theta_{f}}{2}}|0\rangle+e^{i\phi_f}\sin{\frac{\theta_{f}}{2}}|1\rangle$.

ii) When $t\geq{t_f}$, the whole system's Hamiltonian is represented by $\frac{\omega_0}{\cos\theta_f}H_f$  with $H_f=[\cos{\theta_f}\sigma_{z}+\sin{\theta_f}\cos{\phi_f}\sigma_{x}+\sin{\theta_f}\cos{\phi_f}\sigma_{y}]$. It is easy to check that $[H_f,|\psi_f\rangle\langle\psi_f|]=0$.

iii) Since Eq. (\ref{4-0}) holds, $|\omega_0\cdot\tan\theta_f\cdot\cos\phi_f|\leq{g_0}$ and  $|\omega_0\cdot\tan\theta_f\cdot\sin\phi_f|\leq{g_0}$. Note that $g\leq{g_0}$, we conclude that the permissible controls given by Eqs. (\ref{T5-01}-\ref{T5-02}) belongs to $\Omega_{B}(g_0)$.

From the aforementioned observations i)-iii), we conclude that the controlled qubit system  Eq. (\ref{2})  is $P_{\theta_f,\phi_f}$-stabilizable from $\Sigma$ at $t_0$ by $\Omega_{B}(g_0)$ if Eq. (\ref{4-0}) holds.

(2) Second, we need to prove that if the controlled qubit system  Eq. (\ref{2})  is $P_{\theta_f,\phi_f}$-stabilizable from $\Sigma$ at $t_0$ by $\Omega_{B}(g_0)$, then Eq. (\ref{4-0}) holds.

If there exist permissible controls in $\Omega_{B}(g_0)$ such that Eq. (\ref{2})  is $P_{\theta_f,\phi_f}$-stabilizable from $\Sigma$, then there exists static Hamiltonian $H_c=u_x\sigma_{x}+u_y\sigma_{y}\in\Omega_{B}(g_0)$ such that $H_0+H_c$ satisfies $[H_0+H_c,|\psi_f\rangle\langle\psi_f|]=0$.
Note that $[H_0+H_c,|\psi_f\rangle\langle\psi_f|]=0$ if and only if there exists such a scale $\gamma\in{R}$ that $H_0+H_c=\gamma[\cos{\theta_f}\sigma_{z}+\sin{\theta_f}\cos{\phi_f}\sigma_{x}+\sin{\theta_f}\cos{\phi_f}\sigma_{y}]$. Therefore, the following equations should hold simultaneously:
\begin{equation}
\label{4-3z}
\omega_0=\gamma\cos{\theta_f}
\end{equation}
and
\begin{equation}
\label{4-3x}
u_x=\gamma\sin{\theta_f}\cos{\phi_f}
\end{equation}
and
\begin{equation}
\label{4-3y}
u_y=\gamma\sin{\theta_f}\sin{\phi_f}
\end{equation}
Thus  $u_x=\frac{\omega_0}{\cos{\theta_f}}\sin{\theta_f}\cos{\phi_f}$ and $u_y=\frac{\omega_0}{\cos{\theta_f}}\sin{\theta_f}\sin{\phi_f}$.
Recall that $H_c\in\Omega_{B}(g_0)$, i.e.,  $|u_x|\leq{g_0}$ and $|u_y|\leq{g_0}$, we conclude that Eq. (\ref{4-0}) holds.

 This completes the proof of Theorem 1.

 \subsection{Proof of Theorem 2}
 To prove the theorem,  it is sufficient to  show there always exist  bounded time-continuous functional controls in $\Omega_{BC}(g_0)$
 to transfer the qubit system from an arbitrary initial state
$|\psi(t_0)\rangle=|\psi_0\rangle=\cos\frac{\theta_0}{2}|0\rangle+e^{i\phi_0}\sin\frac{\theta_0}{2}|1\rangle\in\Sigma$
to another arbitrary target state
$|\psi(t_{f})\rangle=|\psi_f\rangle=\cos\frac{\theta_f}{2}|0\rangle+e^{i\phi_f}\sin\frac{\theta_f}{2}|1\rangle\in\Sigma$ with $t_0<t_{f}<+\infty$,
and  $|\psi(t)\rangle{\in}C_{\theta_f}$.

 We construct the following permissible controls
\begin{equation}
\label{4-2x}
u_{x}(t)=g(t)\cos\omega_0{(t-t_1)}
\end{equation}
and
\begin{equation}
\label{4-2y}
u_{y}(t)=-g(t)\sin\omega_0{(t-t_1)}
\end{equation}
where
\begin{equation}
\label{4-2gt}
g(t)=\left\{\begin{array}{ll}
0 &t\in[t_0,t_1)\\
g[1-(\frac{t_{1}+t_{f}-2t}{t_{f}-t_{1}})^{n}]& t\in[t_{1},\frac{t_1+t_{f}}{2})\\
g[1-(\frac{2t-t_{1}-t_{f}}{t_{f}-t_{1}})^{n}]& t\in[\frac{t_{1}+t_f}{2},t_{f})\\
0& t\in[t_f,+\infty)
\end{array}\right.
\end{equation}
with
 \begin{equation}
\label{4-1t1}t_1=\frac{\phi_0+\frac{3\pi}{2}}{\omega_0}+t_0
\end{equation}
and
\begin{equation}
\label{4-1k}
k_{dn}=\min\{k\in{Z^+}|k\geq\frac{(n+1)\omega_0}{ng_0}\frac{4\pi+\theta_f-\theta_0}{2\pi}+\frac{\phi_f}{2\pi}-\frac{1}{4}\}
\end{equation}
and
\begin{equation}
\label{4-1g}
g=\omega_0\frac{n+1}{n}\frac{4\pi+\theta_f-\theta_0}{2k_{dn}\pi+\frac{\pi}{2}-\phi_f}\leq{g_0}
\end{equation}
and
\begin{equation}
\label{4-1tf}
t_f=\frac{2k_{dn}\pi+{2}{\pi}-\phi_f+\phi_0}{\omega_{0}}+t_0
\end{equation}

After some calculations, it is demonstrated that  $|\psi(t_{1})\rangle=\cos{\frac{\theta_{0}}{2}}|0\rangle+i\sin{\frac{\theta_{0}}{2}}|1\rangle$.
and $|\psi(t_{f})\rangle=\cos{\frac{\theta_{f}}{2}}|0\rangle+e^{i\phi_f}\sin{\frac{\theta_{f}}{2}}|1\rangle$.
Furthermore we have $|\psi(t)\rangle=\cos{\frac{\theta_{f}}{2}}|0\rangle+e^{i[\phi_f-\omega_0(t-t_f)]}\sin{\frac{\theta_{f}}{2}}|1\rangle\in{C_{\theta_f}}$
when $t\geq{t_f}$.

Because the aforementioned analyses hold for any pair of initial and target states, this implies that one can always construct a local $n^{th}-$order function to dynamically modulate amplitude and further steer quantum systems from an arbitrary initial state to another arbitrary target state.

Therefore, we conclude that   $\forall\theta_f\in{[0,\pi]}$, the controlled qubit system  Eq. (\ref{2})  is $C_{\theta_f}$-stabilizable from $\Sigma$ by $\Omega_{BC}(g_0)$.
 \subsection{Proof of Theorem 3}
\textbf{Proof}: Notice in the proof of Theorem 2 that one can choose
 $t_1=\frac{\phi_0-\frac{\pi}{2}}{\omega_0}+t_0$ and $t_f=\frac{2k_{dn}\pi-\phi_f+\phi_0}{\omega_{0}}+t_0$ if $\phi_0-\frac{\pi}{2}\geq0$,
and observe that one can select
\begin{equation}
\label{theorem2-1}
k_{dn}=\min\{k\in{Z^+}|k\geq\frac{(n+1)\omega_0}{ng_0}\frac{\theta_f-\theta_0}{2\pi}+\frac{\phi_f}{2\pi}-\frac{1}{4}\}
\end{equation}
 and
\begin{equation}
\label{theorem2-2}
g=\omega_0\frac{n+1}{n}\frac{\theta_f-\theta_0}{2k_{dn}\pi+\frac{\pi}{2}-\phi_f}\leq{g_0}
\end{equation}
if $\theta_f-\theta_0\geq0$.

Therefore, we can carry on the analysis to estimate the transition time $t_f-t_0$ by considering four different cases:

Case 1:  $\phi_0-\frac{\pi}{2}<0$ and  $\theta_f-\theta_0<0$.

If  $n>\frac{8\omega_0}{g_0}$, then $\frac{2\omega_0}{ng_0}<\frac{1}{4}$. Furthermore, we have from Eq. (\ref{4-1k}) that  $k_{dn}\leq\frac{2\omega_0}{g_0}+2$.

Thus, if we choose  $n>\frac{8\omega_0}{g_0}$, then
\begin{equation}
\label{4-2a1}
t_f-t_0=\frac{2k_{dn}\pi+{2}{\pi}-\phi_f+\phi_0}{\omega_{0}}\leq{\frac{4\pi}{g_0}+\frac{8\pi}{\omega_0}}.
\end{equation}

Case 2:  $\phi_0-\frac{\pi}{2}<0$ and  $\theta_f-\theta_0\geq0$.

If  $n>\frac{2\omega_0}{g_0}$, then $\frac{\omega_0}{2ng_0}<\frac{1}{4}$. Furthermore we obtain from Eq. (\ref{theorem2-1}) that  $k_{dn}\leq\frac{\omega_0}{2g_0}+2$.

Thus, if we choose  $n>\frac{2\omega_0}{g_0}$, then
\begin{equation}
\label{4-2b2}
t_f-t_0=\frac{2k_{dn}\pi+{2}{\pi}-\phi_f+\phi_0}{\omega_{0}}\leq{\frac{\pi}{g_0}+\frac{8\pi}{\omega_0}}.
\end{equation}

Case 3:  $\phi_0-\frac{\pi}{2}\geq0$ and  $\theta_f-\theta_0<0$.

If  $n>\frac{8\omega_0}{g_0}$, then $\frac{2\omega_0}{ng_0}<\frac{1}{4}$. Furthermore, we have from Eq. (\ref{4-1k}) that $k_{dn}\leq\frac{2\omega_0}{g_0}+2$.

Thus, if we choose  $n>\frac{8\omega_0}{g_0}$, then
 \begin{equation}
\label{4-2c3}
t_f-t_0=\frac{2k_{dn}\pi-\phi_f+\phi_0}{\omega_{0}}\leq{\frac{4\pi}{g_0}+\frac{6\pi}{\omega_0}}.
\end{equation}

Case 4:  $\phi_0-\frac{\pi}{2}\geq0$ and  $\theta_f-\theta_0\geq0$.

If  $n>\frac{2\omega_0}{g_0}$, then $\frac{\omega_0}{2ng_0}<\frac{1}{4}$. Furthermore, we obtain from Eq. (\ref{theorem2-1}) that $k_{dn}\leq\frac{\omega_0}{2g_0}+2$.

Thus,  if we choose  $n\geq\frac{2\omega_0}{g_0}$, then

\begin{equation}
\label{4-2d4}
t_f-t_0=\frac{2k_{dn}\pi-\phi_f+\phi_0}{\omega_{0}}\leq{\frac{\pi}{g_0}+\frac{6\pi}{\omega_0}}
\end{equation}

From Eqs.(\ref{4-2a1}-\ref{4-2d4}),   we have
\begin{equation}
\label{T3-1}
t_f-t_0\leq{\frac{4\pi}{g_0}+\frac{8\pi}{\omega_0}}.
\end{equation}
 for any $|\psi_0\rangle\in{\Sigma}$ and any $|\psi_f\rangle\in{C_{\theta_f}}$.

Therefore, this completes the proof of Theorem 3.

%\bibliographystyle{plain}        % Include this if you use bibtex
%\bibliography{autosam}           % and a bib file to produce the
                                 % bibliography (preferred). The
                                 % correct style is generated by
                                 % Elsevier at the time of printing.

%\appendix
%\section{A summary of Latin grammar}    % Each appendix must have a short title.
%\section{Some Latin vocabulary}         % Sections and subsections are supported
                                        % in the appendices.
\end{document}